\documentclass[conference]{IEEEtran}
\IEEEoverridecommandlockouts

\def\BibTeX{{\rm B\kern-.05em{\sc i\kern-.025em b}\kern-.08em
    T\kern-.1667em\lower.7ex\hbox{E}\kern-.125emX}}

\usepackage{hyperref} 
\usepackage{booktabs} 
\usepackage{caption} 
\usepackage{xspace}
\usepackage{enumitem}
\usepackage{amsmath}
\usepackage{algorithm}
\usepackage{algpseudocode}
\usepackage{svg}
\usepackage{multirow} 
\usepackage{float}
\usepackage{footnote}

\algrenewcommand\algorithmicrequire{\textbf{Input:}}
\algrenewcommand\algorithmicensure{\textbf{Output:}}
\algnewcommand\algorithmicparfor{\textbf{parallel for}}
\algdef{SE}[PARFOR]{ParFor}{EndParFor}[1]
{\algorithmicparfor\ #1\ \algorithmicdo}
{\algorithmicend\ \algorithmicparfor}
\algrenewcommand{\algorithmiccomment}[1]{\hfill // \textit{#1}}
\DeclareMathOperator*{\argmax}{arg\,max}

\newcommand{\sys}[0]{\textsc{EfficientIMM}\xspace}

\title{Sample Paper Title}
\author{
    \IEEEauthorblockN{Hanjiang Wu\IEEEauthorrefmark{3}\IEEEauthorrefmark{1}, 
                      Huan Xu\IEEEauthorrefmark{3}\IEEEauthorrefmark{1}, 
                      Joongun Park\IEEEauthorrefmark{3}, 
                      Jesmin Jahan Tithi\IEEEauthorrefmark{2}, \\
                      Fabio Checconi\IEEEauthorrefmark{2}, 
                      Jordi Wolfson-Pou\IEEEauthorrefmark{2}, 
                      Fabrizio Petrini\IEEEauthorrefmark{2}, 
                      Tushar Krishna\IEEEauthorrefmark{3}}
    \IEEEauthorblockA{\IEEEauthorrefmark{3}Georgia Institute of Technology, USA \hspace{1cm} \IEEEauthorrefmark{2}Intel Corporation}
}

\begin{document}
\title{Enhancing Scalability and Performance in Influence Maximization with Optimized Parallel Processing}
\maketitle

\let\thefootnote\relax\footnotetext{\IEEEauthorrefmark{1}These authors contributed equally to this work.}

\begin{abstract}

Influence Maximization (IM) is vital in viral marketing and biological network analysis for identifying key influencers. Given its NP-hard nature, approximate solutions are employed. This paper addresses scalability challenges in scale-out shared memory system by focusing on the state-of-the-art Influence Maximization via Martingales (IMM) benchmark. To enhance the work efficiency of the current IMM implementation, we propose \sys with key strategies, including new parallelization scheme, NUMA-aware memory usage, dynamic load balancing and fine-grained adaptive data structures.
Benchmarking on a 128-core CPU system with 8 NUMA nodes, \sys demonstrated significant performance improvements, achieving an average 5.9x speedup over Ripples across 8 diverse SNAP datasets, when compared to the best execution times of the original Ripples framework.
Additionally, on the Youtube graph, \sys demonstrates a better memory access pattern with 357.4x reduction in L1+L2 cache misses as compared to Ripples. 
\end{abstract}


\section{Introduction}

The advent of large-scale data and advanced computational frameworks has significantly deepened our exploration of social networks, positioning 
Influence Maximization (IM) as a pivotal research area. This research domain aims to identify the most influential nodes within a network, optimizing the spread of information, behaviors, or products. These insights are invaluable not only for marketing and public health initiatives but also for enhancing our understanding of the complex dynamics of human interaction and information dissemination across vast networks \cite{domingos2001mining, leskovec2007celf, ye2012recomm}.

The IM problem is formally delineated within the framework of graph theory. In brief, the goal of influence maximization is to be capable of identifying a subset of the most influential vertices in an input graph. The challenge inherent to the IM problem arises from its NP-hard nature, indicating the absence of a deterministic polynomial-time algorithm capable of deriving the optimal solution \cite{kempe2003maximizing}. Therefore, it has prompted the development of numerous approximation algorithms. As a sketch-based approximation algorithm, Influence Maximization via Martingales (IMM) is renowned for its robust efficiency and broad adaptability, making it ideal for large-scale social networks \cite{li2018survey}. As first introduced by Tang et al.~\cite{tang2015influence}, IMM employs advanced sampling techniques known as Reverse Influence Sampling (RIS) to generate sketches known as random reverse reachable sets or \textbf{RRRsets} for the input graph. To guarantee the quality of the sampled sketches, IMM utilizes martingale probability theory, which also ensures the linear time-to-solution cost corresponding to the increasing graph size ~\cite{tang2015influence}. Recognizing the growing significance and potential of IMM, it has been identified as a key workflow for further development and research.



Significant efforts have been made to adapt IMM for handling different large-scale data inputs~\cite{shahrouz2021gim,minutoli2019fast,minutoli2020curipples}. We focused on Ripples, a widely adopted open-source IMM framework, particularly suited to the complex and large-scale structure of modern social networks ~\cite{minutoli2019fast}. One of the primary challenges posed by Ripples is algorithm's inability to scale effectively on the latest shared memory server-grade CPUs with multi-NUMA and multi-socket architecture. The system used for evaluation in the original Ripples paper, as described by Minutoli et al. \cite{minutoli2019fast}, consists of a CPU with 10 cores, shared memory and cache spaces, and no NUMA control. Given the trend of shifting to a CPU system with many more NUMA nodes and also the increasing interest of designing scale-out shared memory system, many important benchmarks and frameworks have been adapted into this trend in system \cite{Zhang-numa-aware2015}. Therefore, it is crucial for us to strive for better performance on Ripples targeting the multi-NUMA systems.




With our assessments of the Ripples algorithm across a variety of datasets on a multi-NUMA CPU system, we have pinpointed several critical limitations that hinder its performance: (i) The parallelization strategy employed for RIS sketches demonstrates suboptimal scalability across all CPU workers, (ii) excessive data access coupled with poor data locality, (iii) workload imbalance resulted from several aspects of graph processing.


To address these challenges, this paper first analyzes social graphs and identifies features that impede the scalability of the algorithm. Then, this work introduces a new parallelization strategy that improves scalability far beyond what the current Ripples framework achieves. In addition to the parallelization strategy, we also redesigned the IMM algorithm for NUMA-awareness and employed optimizations based on fundamental characteristics of social graphs.

This paper makes several contributions to the field of IMM, outlined as follows:
\begin{itemize}
\item We conduct a comprehensive profiling of the current state-of-the-art solutions for the IMM problem with various contexts of the social graphs, pinpointing critical bottlenecks that necessitate optimization.
\item We introduce a new shared-memory partitioning and distribution strategy that greatly enhances the efficiency of constructing the influence counts for each vertices based on the sampled RIS sketches through concurrent updates. Using the new parallelization technique, we further adapted the algorithm with NUMA-aware data structures and greatly improved data reuse.
\item We have developed optimization techniques to address workload imbalances arising from the varying attributes of different graphs, including adaptive data structures and dynamic job balancing strategies. 
\item We applied these optimizations to eight selected SNAP datasets, achieving an impressive performance speedup ranging from 1.6 to 12.1 times compare to Ripples' best runtime. Also, our work efficient algorithm also brings 22.4x to 357.4x less L1+L2 cache misses as compared to Ripples.
\end{itemize}

\section{Background}

\subsection{Influence Maximization}

Consider a directed graph \(G = (V, E)\), where \(V\) represents the set of vertices in the network, and \(E \subseteq V \times V\) represents the directed edges indicating the influence relationships between the vertices.

The goal of IM is to identify a subset of vertices \(S \subseteq V\), subject to \(|S| = k\), that maximizes the expected influence spread $\sigma(S)$ in the network, under a given influence diffusion model  \(M\). 
\(k\) is a predefined budget limiting the size of \(S\), and $\sigma(S)$ quantifies the expected number of vertices influenced through the diffusion process originating from \(S\).

The influence diffusion process is typically conceptualized through two prevalent models known in contemporary research.

\noindent
\textbf{Independent Cascade (IC) Model:} Initially, all vertices in \(S\) are activated, and the rest are inactive. At each timestep, each newly activated vertex \(u\) has a single chance to activate each of its currently inactive neighbors \(v\) with a probability \(p_{uv}\), where \(p_{uv} \in [0, 1]\) is associated with the edge \((u, v) \in E\). The process continues until no new activation occurs.

\noindent
\textbf{Linear Threshold (LT) Model:} Each vertex \(v \in V\) is assigned a threshold \(T_v\) uniformly at random from the interval \([0, 1]\). Each edge \((u, v) \in E\) has a weight \(w_{uv}\), with the constraint \(\sum_{u:(u,v) \in E} w_{uv} \leq 1\) for each \(v\).
A vertex \(v\) becomes activated if the sum of the weights from its activated neighbors exceeds its threshold, i.e., 
\[
\textstyle\sum_{u \in \text{ActivatedNeighbors}(v)} w_{uv} \geq T_v
\]
The diffusion starts with the activation of vertices in \(S\) and proceeds in discrete steps until no further activations can occur.
The IM problem is NP-hard for both the IC and LT models. However, due to the submodularity of the influence spread function ($\sigma(S)$), a greedy approach guarantees a solution within \((1 - 1/e)\) of the optimal, where \(e\) is the base of the natural logarithm.



\subsection{Parallelized IMM Algorithm} \label{sec:par_IMM}
The IMM algorithm marks a significant advancement in solving the IM problem by offering a near-optimal solution with reduced computational complexity. Despite its efficiency, the exponential growth in the size and complexity of social networks necessitates further enhancements to IMM’s scalability and performance. Parallelizing IMM, which involves adapting the algorithm to execute concurrently across multiple processing units, emerges as a crucial strategy to meet these challenges. Previous studies have investigated this problem~\cite{minutoli2019fast,minutoli2020curipples}.



The workflow of the IMM algorithm, detailed in Algorithm \autoref{alg:IMM_workflow}, is structured into two principal phases: Sampling and Selection. 
In the Sampling phase, the algorithm iteratively generates RRR sets and identifies the most influential nodes. Since IMM provides an approximate solution, it requires a user-specified tunable hyperparameter, $\epsilon$. A smaller value of $\epsilon$ results in a more representative sampling subset of the input graph. This iterative sampling process is crucial for determining the convergence parameter, $\theta$, ensuring it meets the specified approximation accuracy. In Algorithm \autoref{alg:IMM_workflow}, we symbolize Theta\_Estimation, OPT\_Lower\_Bound and Set\_Theta to represent the steps that determine the iterative process of $\theta$ generation as mentioned in Tang et al. \cite{tang2015influence}. 
Subsequently, the Selection phase involves repeating these methods to refine and finalize the selection of influencers. 
Two key kernels in both phases, \textit{Generate\_RRRsets} and \textit{Find\_Most\_Influential\_Set}, are critical for performance estimation of the IMM algorithm.
Therefore, Ripples focuses on the parallel execution techniques for these two core kernels, as introduced below.

\begin{algorithm}
\caption{IMM Workflow \cite{minutoli2019fast,tang2015influence}}
\label{alg:IMM_workflow}
\begin{algorithmic}[1]
\Require Graph $G$, target \# of seeds $k$, quality approximation \(\epsilon\)
\Ensure \(S\) 
\Statex \textit{// Sampling Phase:}
\For{\(i = 1 \rightarrow log{|V|}\)}
    \State \(\theta' \leftarrow \text{Theta\_Estimation}(G, k, i, \epsilon)\)
    \State \(RRRsets \leftarrow \textbf{Generate\_RRRsets}(G, \theta', RRRsets)\)  
    \State \(S\_tmp \leftarrow \textbf{Find\_Most\_Influential\_Set}(G, RRRsets)\)
    \State \(LB \leftarrow \text{OPT\_Lower\_Bound}(G, i, S\_tmp)\)
\EndFor
\State \(\theta \leftarrow \text{Set\_Theta}(G, k, \epsilon, LB)\)
\If{\(\theta' < \theta\)} \Comment{Need to generate more RRRsets}
    \State \(RRRsets \leftarrow \textbf{Generate\_RRRsets}(G, \theta, RRRsets)\) 
\EndIf
\Statex \textit{// Selection Phase:}
\State \(S \leftarrow \textbf{Find\_Most\_Influential\_Set}(G, RRRsets)\)
\State \Return{\(S\)}
\end{algorithmic}
\end{algorithm}

\noindent
\textbf{Generate\_RRRsets: }
The \textit{Generate\_RRRsets} algorithm generates a $\theta$ number RRRsets using through probabilistic Breadth-First Search (BFS).
With a total number of p threads, each thread works on $\theta \over p$ RRRsets in parallel. 
In the probabilistic BFS algorithm, the starting node for doing the BFS is uniformly selected from all the nodes in the graph. Under the context of RIS, the initial nodes are activated by default and placed in the returned RRRset and a queue used to track activated but not yet traversed nodes. Then, the generation of each RRRset is done by adding more nodes into the RRRset depending on the aforementioned diffusion model. This process will stop when no more nodes are activated. The detailed algorithm is illustrated along with our design in \autoref{sec:proposed_algo}.


\noindent
\textbf{Find\_Most\_Influential\_Set: }
\textit{Find\_Most\_Influential\_Set} is another key kernel in the parallel IMM algorithm (as highlighted in Algorithm \autoref{alg:IMM_workflow}). After the generation of RRRsets, we begin identifying the nodes (seeds) that have the greatest impact on these RRRsets. In theory, the selected seeds will have the most significant influence on our sampled graph and are mathematically ensured to be representative of the original input graph~\cite{tang2015influence}.

 The algorithm starts off by assigning each thread with a number of vertices, so the work per thread here is $|V| \over p$. Then, a vector of $|V| \over p$ counters is created to note the frequency of all the nodes showing up in the RRRsets for each thread. When updating the counters, each thread needs to traverse through all the RRRsets and check if it contains the vertices to be updated in the counter. After the initial updates with respective to all the RRRsets generated, the highest counted vertex is reduced from all the thread-local vertex counters. With the selected seed, we do a decoupling step to remove the selected seed's impact in the corresponding RRRsets. This step is an iteration process to decrement those vertex counters if they in the same RRRsets as the selected seed. At last, we repeat the selection of seed for another $k-1$ times to find the remaining most influential seeds and finalize the most influential set to return.
\section{Motivation}\label{sec:motivation}



\subsection{Characterizing Graphs within the IMM}
Broder et al. \cite{journals/BroderKMRRSTW00} identified a fundamental property of graph structures in the web: most nodes form a single connected component if links are treated as undirected edges, called the "strongly connected component" (SCC). This property of the graph has significant performance implications for IMM. In \textit{Generate\_RRRsets} kernel, it performs $\theta$ RRRsets via probabilistic BFS with each RRRset having the potential to cover more than 80\% of the nodes in the graph due to SCC. This leads to increased memory footprint and computational overhead. In addition, in \textit{Find\_Most\_Influential\_Set} kernel, it selects the most influential nodes among the constructed RRRsets, in which the performance can be highly impacted by the size of a large RRRset. The SCC can be embodied at different degrees in relation to the diffusion model used. In the IC model, the influence is spread with independent probability, while in the LT model, the influence is spread with the impact contributed by all the neighboring nodes. Therefore, one of the most salient characteristics is that IC has a higher chance of forming an SCC because all the nodes have a chance to activate an node. Therefore, under the IC model, the average number of vertices in RRRsets is large, but \(\theta\), the number of RRRsets required for accurate influence estimation, is relatively small, usually in the 1e3 to 1e4 range. Conversely, under the LT model, the average number of vertices in RRRsets is smaller than IC, but \(\theta\) is larger, usually in the 1e8 to 1e9 range.




\subsection{Performance Analysis : Ripples}

\begin{figure}[htbp]
\centerline{\includegraphics[width=8.5cm]{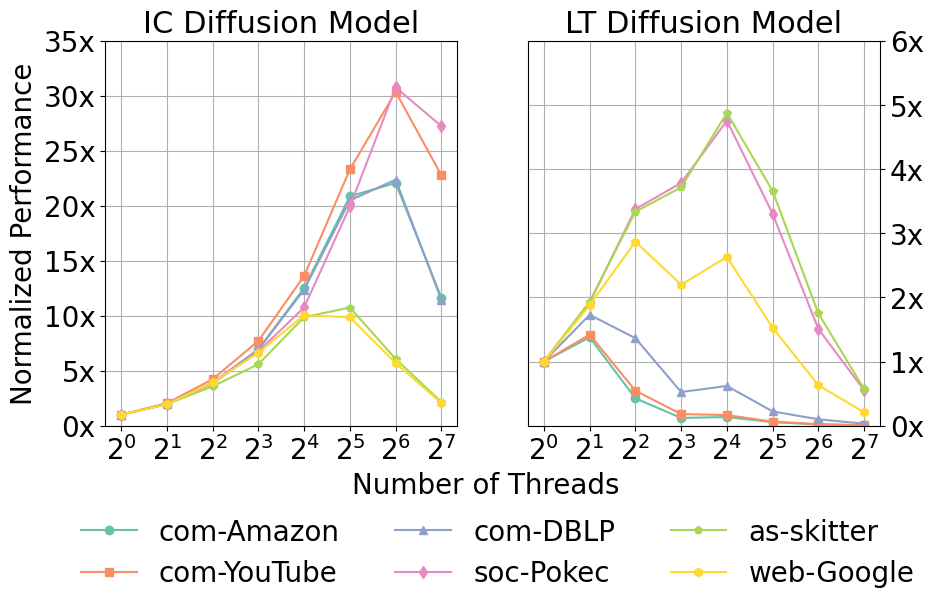}}
\caption{Ripples Strong Scaling Performance}
\label{fig:strong_scaling}
\end{figure}

The current implementation of Ripples on shared memory system has limitations stemming from the fundamental graph properties discussed above. Figure~\ref{fig:strong_scaling} displays the experimental results for the strong scaling of Ripples as the number of threads increases. 
The results reveal that Ripples encounters scalability limits in both the LT model and the IC models. Specifically, scalability in Ripples is limited after 4 threads in the LT model and 32 threads in the IC model. These indicate that although Ripples can efficiently handle the workload at the beginning, its ability to scale quickly diminishes beyond certain thread counts. 


\begin{figure}[htbp]
\centerline{\includegraphics[width=8.5cm]{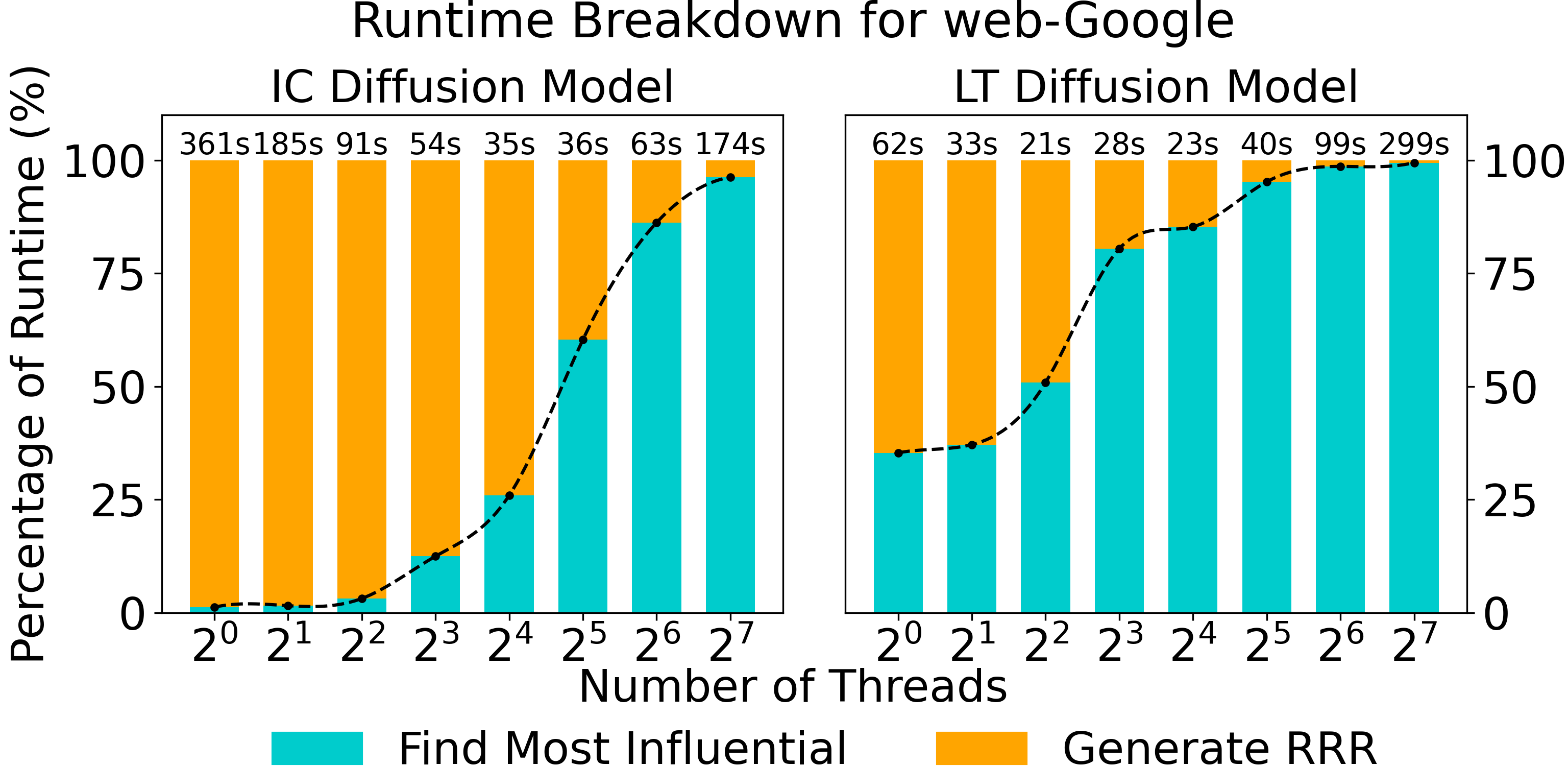}}
\caption{Ripples Runtime Breakdown}
\label{fig:runtime_breakdown}
\end{figure}


Furthermore, the breakdown results shown in Figure~\ref{fig:runtime_breakdown} show that  Independent Cascade (IC) and Linear Threshold (LT) models using 1 to 128 cores on the web-Google dataset from the SNAP collection. The trends are consistent across all SNAP datasets. The analysis reveals that \textit{Generate\_RRRsets} and \textit{Find\_Most\_Influential\_Set} dominate the runtime, with \textit{Find\_Most\_Influential\_Set} significantly impacting scalability as the number of cores increases. Notably, on the web-Google dataset, Ripples fails to scale beyond 16 physical cores on a 128 physical core system for both the LT and IC diffusion models. 

This limitation is primarily due to \textit{Find\_Most\_Influential\_\allowbreak Set}, which experiences a significant slowdown when utilizing more than 15 threads and performs ineffective computations. The scalability challenges of IMM and performance bottlenecks associated with Ripples will be explored in more detail in the following section. 

\subsection{Challenges}


\begin{table}[htbp]
\label{tab:rrrset_characteristics}
\caption{Input Graph and Ripples RRRset Characteristics (with IC diffusion model; and input hyperparam of $\epsilon=0.5$)}
\vspace{-0.3cm}
\begin{center}
\begin{tabular}{|p{1.6cm}|p{1.2cm}|p{1.6cm}|p{1.2cm}|p{1.2cm}|}
\hline
\textbf{Graph} & \textbf{Nodes} & \textbf{Edges} & \textbf{Average RRRset Coverage} & \textbf{Max RRRset Coverage} \\
\hline
com-Amazon & 334,863 & 925,872 & 61.3\% & 79.6\% \\
\hline
com-YouTube & 1,134,890 & 2,987,624 & 32.7\% & 59.9\% \\
\hline
com-DBLP & 317,080 & 1,049,866 & 51.4\% & 78.9\% \\
\hline
com-LJ & 3,997,962 & 34,681,189 & 68.0\% & 84.1\% \\
\hline
soc-Pokec & 1,632,803 & 30,622,564 &  60.1\% & 78.5\% \\
\hline
as-Skitter & 1,696,415 & 11,095,298 & 1.6\% & 5.4\% \\
\hline
web-Google & 875,713 & 5,105,039 & 17.4\% & 54.8\% \\
\hline
Twitter7 & 41,652,230 & 1,468,365,182 & 59.8\% & 88.0\% \\
\hline
\end{tabular}
\label{tab:merged_graph_rrrset}
\end{center}
\end{table}

\noindent
\textbf{Challenge 1: Memory Traversal: }Our analysis highlights that limited scalability in Ripples predominantly stems from extensive memory traversal during the \textit{Find\_Most\_Influential\_Set}. 
A key factor is the counting of vertex occurrences within RRRsets. 
During this operation, each thread independently accesses all RRRsets to determine if an RRRset contains the specific vertices tallied for that thread.
While this method partitions the workload among all the workers independently, it significantly increases memory, thus exacerbating scalability challenges. 
A similar issue arises during the counters updates, after selecting the most influential seed vertex $v$. Each RRRset containing $v$ must be updated: the set is removed, and the occurrence counters for all vertices within that set are decremented. This process again requires all threads to repeatedly access the RRRsets until the most influential set of $k$ vertices is determined. 

\noindent
\textbf{Challenge 2: Large RRRsets due to SCC: } The presence of strongly connected components (SCCs) in graphs poses a significant challenge for the efficient generation and processing of RRRsets in Ripples. As shown in Table \ref{tab:merged_graph_rrrset}, the max RRRset coverage rate, which represents the fraction of vertices covered by each RRRset, is over 54\% for all datasets except as-Skitter, which is a road network. This high coverage rate is a direct consequence of the SCC structure, where a large portion of the graph is interconnected, and a single randomized BFS potentially reach a substantial fraction of the vertices. The generation of dense RRRsets due to SCCs leads to several performance bottlenecks in Ripples. Firstly, Ripples employs a sorting mechanism for all RRRsets during the \textit{Generate\_RRRsets}, which becomes increasingly inefficient as the RRRset size grows. Secondly, the \textit{Find\_Most\_Influential\_Set} utilizes binary search when updating the global influence counter, which also suffers from the increased RRRset sizes. 
These operations contribute to the scalability limitations of Ripples, particularly when dealing with large-scale graphs containing significant SCCs. Thus, we need to be scrutinized with the selection of specific data structures while we are generating the RRRsets to avoid the unnecessary NUMA access inefficiencies. 

In the following section, we propose our new algorithms that optimize the memory traversal for \textit{Find\_Most\_Influential\_Set} and in the \textit{Generate\_RRRsets} kernel, we will also present an optimized NUMA-aware design and the corresponding algorithm adaptation.


\section{Proposed Algorithm} \label{sec:proposed_algo}

To address the exposed scaling and performance challenges, our work aims to enhance the performance of IMM by refining the existing design and tackling the identified bottlenecks. We propose solutions that optimize memory accesses, balance computational loads, and reduce synchronization dependencies to improve both efficiency and scalability.

\subsection{Partitioning/Parallelization Strategy}

\sys introduces a new strategy for partitioning both vertices and RRRsets, thereby significantly enhancing the algorithm's scalability. Traditional approach requires each thread to go through every RRRset during the \textit{Find\_Most\_Influential\_Set} phase, leading to an increased number of load/store operations. The amount of operations not only escalates with the number of threads but also with the volume of RRRsets, affecting efficiency and scalability. 

\begin{algorithm}
\caption{\sys's Find\_Most\_Influential\_Set}
\label{alg:our_find_most_inf}
\begin{algorithmic}[1]
\Require Graph $G$, target \# of seeds $k$, $RRRsets$
\Ensure \( S \)
\State  $\text{counter} \gets 0$ \Comment{Initialize Global Counter}
\ParFor{\textbf{each} \( R \in RRRsets \)}
    \For{\textbf{each} \( n \in R\)}
    \State \textbf{Atomic} \{
        $\text{counter}[n] \gets \text{counter}[n] + 1$
    \}
    \EndFor
\EndParFor
\State \( S \leftarrow \emptyset \)
\While{\( |S| < k \)}
    \State \(v \gets \textbf{PARALLEL\_REDUCTION}(\text{counter})\) 
    \State \( S \leftarrow S \cup \{v\} \)
    \ParFor{\textbf{each} \( R \in RRRsets \)}
    \If{\(v \in R\)}
    \For{\textbf{each} \(n \in R\)}
        
            \State \textbf{Atomic} \{
                $\text{counter}[n] \gets \text{counter}[n] - 1$
            \}
    \EndFor
    \EndIf
    \State \(\text{RRRsets} \setminus \{R\}\)
    \EndParFor
\EndWhile
\State \Return \( S \)
\end{algorithmic}
\end{algorithm}

\noindent
\textbf{RRRsets Partitioning: }
In contrast to the conventional approach, \sys advocates for the partitioning of RRRsets during the \textit{Find\_Most\_Influential\_Set}  phase. This partitioning ensures that each thread is only accountable for a specific subset of  (as shown in algorithm \autoref{alg:our_find_most_inf}). As a result, a thread in \sys handles only \(\frac{\theta}{p}\) RRRsets, thereby diminishing the memory access by a factor of \(1/p\). This strategic reduction lowers the computational demand and the memory footprint per processor because each thread does not have to traverse all the RRRsets. 



\begin{figure}[htbp]
  \centering
  \includegraphics[width=1.09\columnwidth]{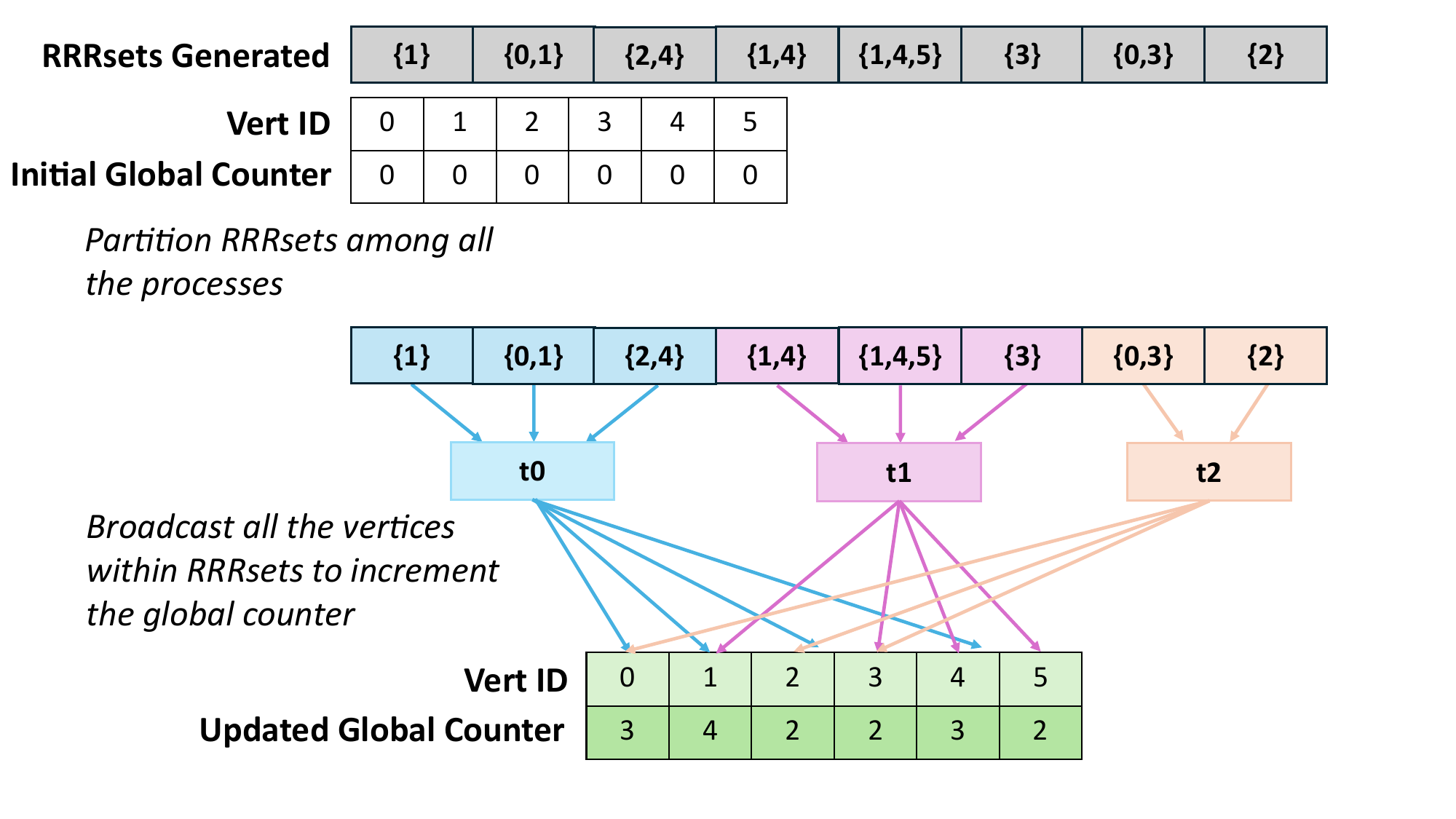}
  \vspace{-0.8cm}
  \caption{Counter updates pattern in our proposed algorithm.}
  \label{fig:our_infmax}
\end{figure}

\noindent
\textbf{Concurrent Updates to the Global Counter:}
As depicted in Figure \ref{fig:our_infmax}, it shows our proposed parallel pattern for counter updates. 
With the partitioned RRRsets, each thread broadcasts to increment the counters of vertices inside the RRRsets atomically. 
A trade-off encountered by \sys involves its requirement for frequent access to shared memory for the global counter, potentially impeding scalability. 
To mitigate this issue, \sys leverages fine-grained memory locking facilitated by indigenous \texttt{C++ atomic add} operation, targeting only a 64-bit region of memory at a time. 
The assembly instruction \texttt{lock incq (\%rbx, \%rdx, 8)} confines the locked memory to a 64-bit quadword, thus preventing the locking of unaffected counters and reducing unnecessary contention overhead. 
This approach minimizes memory congestion and optimizes memory access, improving the overall efficiency of the counter updates. 
Algorithm \autoref{alg:our_find_most_inf} shows the pseudocode our proposed \textit{Find\_Most\_Influential\_Set} algorithm with the new parallelization strategy and the atomic counter updates. The new counter updates pattern is applied both in counting initial vertex occurrences and in adjusting the counter after reducing the vertex with the highest count.

\noindent


\begin{algorithm}
\caption{Generate\_RRRsets with kernel fusion}
\label{alg:bfs_diffusion}
\begin{algorithmic}[1]
\Require Graph $G$, starting vertex $s$, global vertex $counter$, diffusion function $D$
\Ensure Set of activated nodes $RRRset$
\State $Q \leftarrow$ empty queue
\State $RRRset \leftarrow \{s\}$ \Comment{initialize activated set with starting vertex}
\State $visited[s] = true$ \Comment{initialize the visited Bitmap}
\State Enqueue $Q, s$
\While{$Q$ is not empty}
    \State $u \leftarrow$ Dequeue $Q$
    \For{each neighbor $v$ of $u$}
        \If{$!visited[v]$ and D(random($u$,$v$))}
            \State $RRRset \leftarrow RRRset \cup \{v\}$
            \State Enqueue $Q, v$
        \EndIf
    \EndFor
\EndWhile
\Statex \textit{//In-place counter updates.}
\For{\textbf{each} \( n \in RRRset\)}
\State \textbf{Atomic} \{
    $counter[n] \gets counter[n] + 1$
\}
\EndFor

\State \Return $RRRset$
\end{algorithmic}
\end{algorithm}

\noindent
\textbf{Parallel Reduction:} 
With the constructed global counter, we can perform an efficient and scalable parallel reduction to identify the vertex with the highest count (line 9 in Algorithm \autoref{alg:our_find_most_inf}). The parallel reduction is divided into two steps: 1. Each thread finds the regional maxima in its portion of the vertex counters. 2. Traverse all the regional maxima to find the global maxima. 
In the first step, all threads are assigned with a range of consecutive vertices $\{v_{l}, \ldots, v_{h}\}$. 
For each thread, the regional maxima are determined by:
\[
v_{\text{reg}} \leftarrow \argmax_{n \in \{v_{l}, \ldots, v_h\}} (\text{counter}[n])
\]
Subsequently, using a list of regional maxima, the global maximum is reduced by:
\[
v_{\text{glob}} \leftarrow \argmax_{t \in \{1, \ldots, p\}} (\text{regional\_maxima}[t])
\]

\subsection{NUMA-Aware Design and Algorithm Adaptation}

\textbf{NUMA-Aware Data structure: }
Due to the nature of strongly connected components (SCC), memory accesses to certain sets of vertices and edges are highly frequent. This often results in intensive access to a single NUMA node, causing imbalance in bandwidth contention and workload. To address this issue, \sys utilizes NUMA interleaving, distributing vertices and edges across different NUMA nodes to prevent concentrating workload on a single node. This approach effectively mitigates workload imbalance and ensures balanced memory access across threads.
However, while NUMA interleaving alleviates workload imbalance, it increases inter-socket memory access, which can partially offset the benefits of reduced memory contention.

\begin{figure}[htbp]
  \centering
  \includegraphics[width=1\columnwidth]{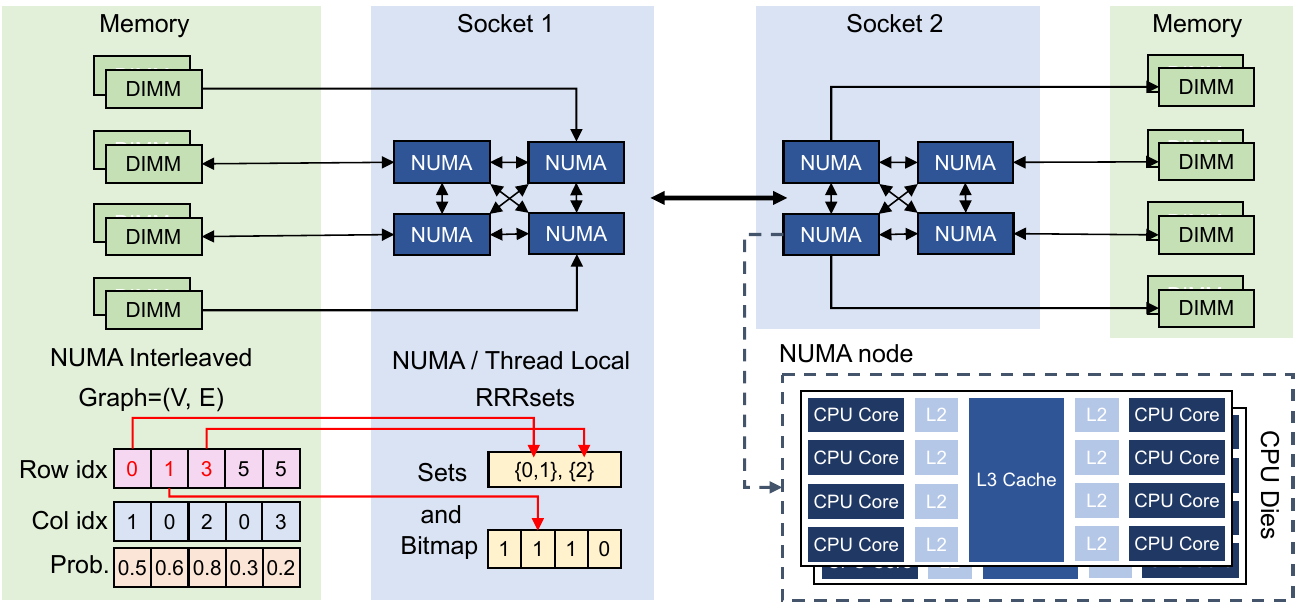}
  \vspace{-0.4cm}
  \caption{System architecure and NUMA-aware data structures}
  \label{fig:numa-aware-arch}
\end{figure}

Profiling the most time-consuming memory accesses revealed that over 30\% of the performance overhead in \textit{Generate\_RRRSets} occurs during BFS, especially when checking the bitmap that stores visited vertices (line 8 of Algorithm~\ref{alg:bfs_diffusion}). 
To reduce this overhead, \sys caches key data structures such as RRR sets and bitmaps to place them closer to the processor. It also leverages \textit{mbind} to ensure each NUMA node operates within its local memory as shown in \ref{fig:numa-aware-arch}. 
This strategy reduces memory access latency and minimizes contention.
Table \ref{tab:numa-datastructure} shows the total effect of using NUMA-aware data structure and the relative performance improvement. 
We observe that with careful data placement, the percentage of core time spent on checking the bitmap improves across all datasets, with the improvement ranging from 38\% to 63\% across the five tested input graphs.

\begin{table}[h!]
\centering
\begin{tabular}{|c|c|c|c|}
\hline
\textbf{Graph} & \textbf{Original} & \textbf{NUMA-aware} & \textbf{Percentage} \\
& \textbf{Data Structure} & \textbf{Data Structure} & \textbf{Improvement} \\
\hline
Amazon & 38.2\% & 23.8\% & 38\% \\ 
\hline
Youtube & 38.6\% & 23.9\% & 38\% \\
\hline
Pokec & 44.9\% & 16.6\% & 63\% \\
\hline
LiveJournal & 46.3\% & 18.5\% & 60\% \\
\hline
Google & 29.0\% & 13.6\% & 53\% \\
\hline
\end{tabular}
\caption{Comparison of Total Core Time Percentage Between Original and NUMA-aware Data Structure with 8 NUMA Nodes Used}
  \label{tab:numa-datastructure}
\end{table}

\textbf{Kernel Fusion:}
In the original Ripples Algorithm, in which, \textit{Generate\_RRRsets} and \textit{Find\_Most\_Influential\_Set} are distinct kernels. Typically, the RRRsets generated from the \textit{Generate\_RRRsets} kernel need to be gathered before moving on to the \textit{Find\_Most\_Influential\_Set} step. The redistribution of all RRRsets to feed into the \textit{Find\_the\_Most\_Influential\_Set} kernel involves unnecessary memory traffic.

 By leveraging our partitioning strategy, we can do \textit{In-Place Data Processing}.  The core concept involves merging the \textit{Generate\_RRRsets} and \textit{Find\_the\_Most\_Influential\_Set} operations into a unified kernel. This fusion strategy saves the load and store counts and enhances data locality. In the following evaluation section, we display the beneficial cache accesses brought by the parallelization strategy and kernel fusion. Algorithm\autoref{alg:bfs_diffusion} is the \textit{Generate\_RRRsets} kernel after the fusion. In Figure \ref{fig:numa-aware-arch}, we exemplify how the kernel fusion facilitates the local generation of the RRRsets, and how it can be reused in combination with counter updates. The red colored row indices of the interleaved graph are ids (0, 1, 2 in this case) of the vertices that are initially activated. After the probabilistic BFS, we can see the RRRSets will be first stored in the cache of the local NUMA nodes. Immediately after the generating RRRsets, we updates vertex counters that match the vertices within each RRRSet.






\subsection{Key Optimizations}



\noindent
\textbf{Adaptive RRRset Representation:} 
Searching within RRRsets is a critical operation that frequently occurs during the seed selection phase. 
Typically, RRRsets are sorted at the end of their generation to facilitate $O(\log n)$ search times for vertex occurrences. 
However, our analysis of datasets exhibiting significant skewness shows that some vertices disproportionately influence others, leading to about 80\% of the sorted RRRsets being discarded in the initial selection round. 
Consequently, the sorting effort, which has a computational cost of $O(n \log n)$, becomes largely inefficient.
Moreover, we observed that the Ripples framework often encounters out-of-memory issues due to the large memory footprint required for storing RRRsets during the sampling step. 
Prior effort to address this challenge has adopted Huffman coding or bitmap coding to compress RRRsets~\cite{chen2022hbmax}. 
While effective in reducing storage requirements, these methods come with a trade-off, notably increasing the computational overhead associated with encoding and decoding.
To address these inefficiency, \sys employs an adaptive representation of RRRsets, alternating between bitmap and a set of vertices based on individual RRRSet's characteristics. 
This design optimizes the storage of RRRsets under different context. 
Based on the empirical analysis, \sys determines the threshold to dynamically switch between different type of the storage. 
Below the threshold, vertex indices are returned as RRRsets, while above the threshold, they are stored as bitmaps.
This adaptive technique ensures both optimal memory storage and efficient computational cost.  
For large datasets like Twitter7 shown in Table \ref{tab:merged_graph_rrrset}, storing all the RRRsets as bitmap with over 40 million nodes will be memory consuming. 
With an adaptive representation, it prunes the RRRsets that are relatively small in size and only use bitmap representations for dense RRRsets. 
This allows $O(1)$ search for dense RRRsets and the $O(log(n))$ cost for sorting those small RRRsets is also cheap and worthwhile.

\begin{figure}[htbp]
  \centering
  \includegraphics[width=8cm]{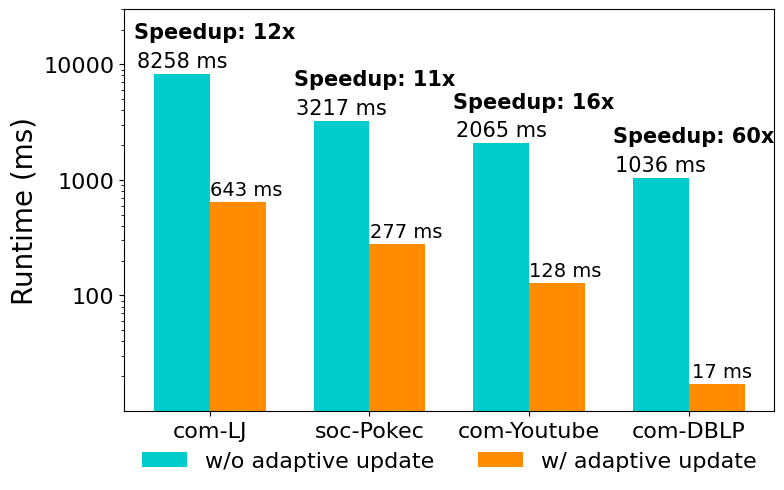}
  \captionsetup{skip=1pt}
  \caption{Runtime Comparison w/ and w/o Adaptive Update when using 128 cores}
  \label{fig:adaptive_runtime}
\end{figure}

\noindent
\textbf{Adaptive Vertex Occurrence Counter Update:}
More optimizations can be done in the seed selection step, when \sys identifies RRRsets containing the selected seeds. It decrements the counters for all vertices in those RRRsets. 
This issue is exacerbated in datasets with severe skewness, where most RRRsets contain the selected seeds. 
These RRRsets, associated with the most influential vertices, tend to be large (due to the SCC effect mentioned in section \ref{sec:motivation}), compounding the inefficiency caused by futile work and extra atomics contention. 
To address this, \sys adaptively updates the global counter rather than merely decrementing vertex occurrences in RRRsets. 
Instead of reducing counts in every identified RRRset, \sys rebuilds the global counter using only those RRRsets that do not contain the selected seeds. 
This method is particularly effective given the number and size of RRRsets involved. 
We observed that this optimization significantly decreases the time required for seed selection. Figure \ref{fig:adaptive_runtime} illustrates the runtime comparison w/ and w/o adaptive counter updates under 128 cores. The results indicate that adaptive counter updates provide a relative speedup ranging from 11.6x to 60.9x across 4 datasets.


\noindent
\textbf{Dynamic Job Balancing:}
The partitioning strategies may influence the distribution of vertex occurrences in RRRsets, as well as the size of each RRRset. This can be embodied at the Algorithm \autoref{alg:our_find_most_inf} that does \textit{Generate\_RRRsets} with kernel fusion and the process of adjusting the counters after finding the most influential seed in Algorithm \autoref{alg:our_find_most_inf}. To prevent workload imbalances across threads, \sys implements dynamic job balancing using a producer-consumer model. When a thread completes all its assigned jobs, it first checks its own work-queue. This approach effectively reduces workload imbalances caused by the significant variance in RRRset sizes, while still preserving the advantages of locality inherent in RRRset/vertex partitioning within each job batch.


\section{Evaluation}
\label{sec: evaluation}


\subsection{Experimental Setup}
Our experimental evaluations were carried out on Perlmutter ~\cite{nersc_perlmutter} with dual-socket AMD EPYC 7763 (2.45GHz) 64-core CPUs, 512 GB of DDR4 memory in total, and 4 NUMA domains per socket. 
The experiment was conducted on 1-128 physical cores mapping on two sockets that ensures both \sys and Ripples do not use hyper-threading. 
For compiler, we used the GCC version 8.1, employing the “-O3” flag. To efficiently conduct experiments under various NUMA configurations, we utilize \textit{numactl} to interleave memory storage across NUMA domains.


\noindent
\textbf{Testing Benchmark:}
Our experimental framework is built upon the latest stable Ripples release (commit: 0a9f3e7) to ensure a fair comparison. 
We utilized OpenMP to facilitate parallel computing in the shared memory, enabling the efficient execution of computations across multiple threads. 
\sys supports two diffusion models: LT and IC models, both used in later evaluation.


\begin{table}[h]
\label{tab:perf IC + LT}
\caption{\sys's Best Runtime Performance (sec)}
\vspace{-0.3cm}
\begin{center}
\scriptsize 
\begin{tabular}{|l|p{1.4cm}|p{1.4cm}|p{1.4cm}|p{1.4cm}|}
\hline
\multirow{2}{*}{\textbf{Graph}} & \multicolumn{2}{c|}{\textbf{Independent Cascade (IC)}} & \multicolumn{2}{c|}{\textbf{Linear Threshold (LT)}} \\
\cline{2-5}
 & \textbf{Ripples} & \textbf{EfficientIMM} & \textbf{Ripples} & \textbf{EfficientIMM} \\
\hline
com-Amazon & 7.93 & 0.97 & 0.93 & 0.16 \\
com-DBLP & 7.10 & 0.94 & 4.2 & 0.85 \\
com-YouTube & 14.07 & 3.0 & 1.23 & 0.14 \\
as-Skitter & 2.3 & 0.45 & 38.96 & 10.59 \\
web-Google & 36.04 & 4.82 & 21.93 & 3.7 \\
soc-Pokec & 59.90 & 36.97 & 40.57 & 10.7 \\
com-LJ & 167.4  & 134.00 & 1.58 & 0.13 \\
twitter7 & OOM & 1645.58 & 2354.7 & 1734.9 \\
\hline
\end{tabular}
\label{table:perf_comparison}
\end{center}
\end{table}

\begin{figure}[h]
  \centering
  \includegraphics[width=9cm]{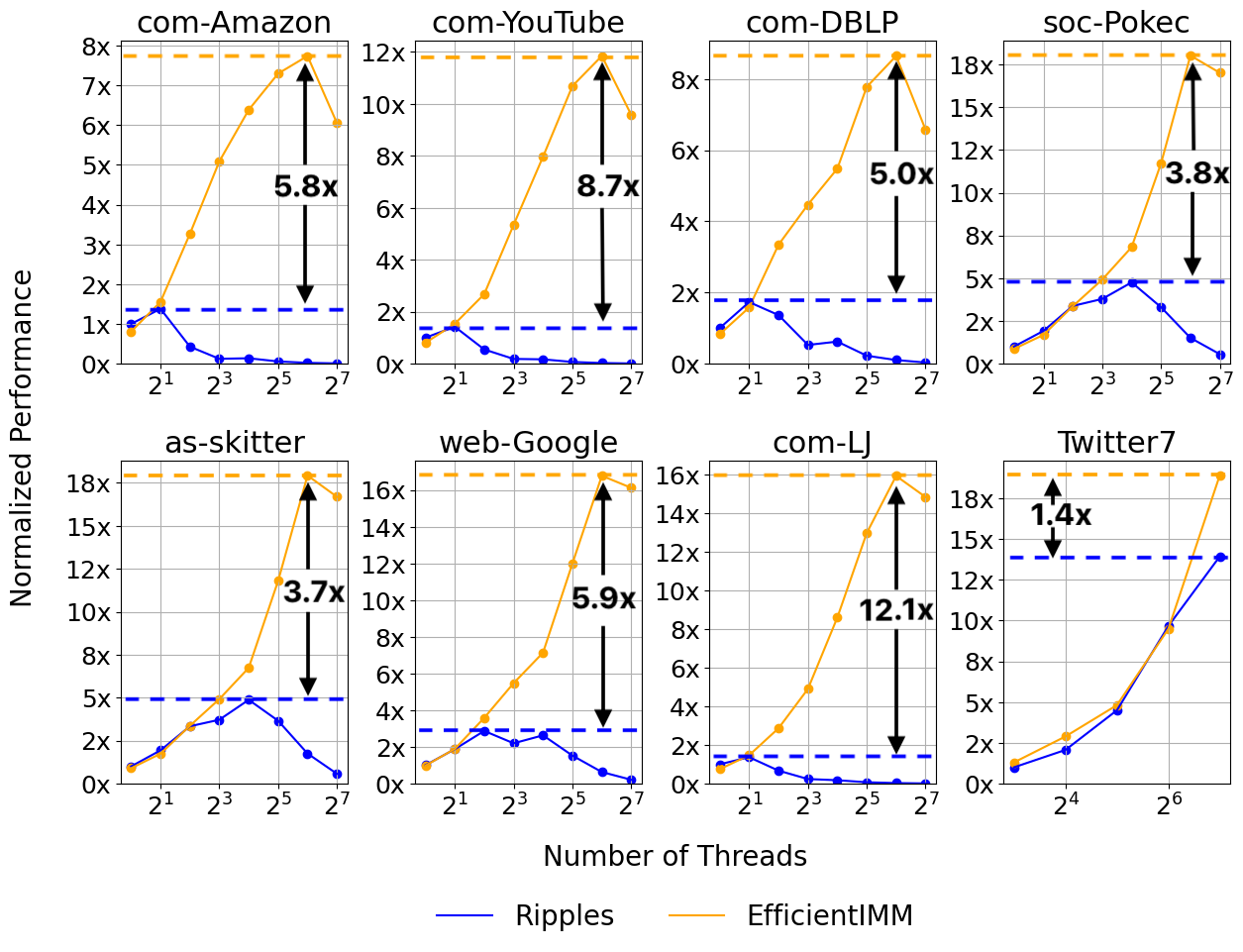}
  \caption{Strong Scaling Performance Normalized to 1 and 8 threads Ripples; LT Diffusion Model; k = 50, $\epsilon$ = 0.5}
  \label{fig:scaling_lt}
\end{figure}

\begin{figure}[h]
  \centering
  \includegraphics[width=9cm]{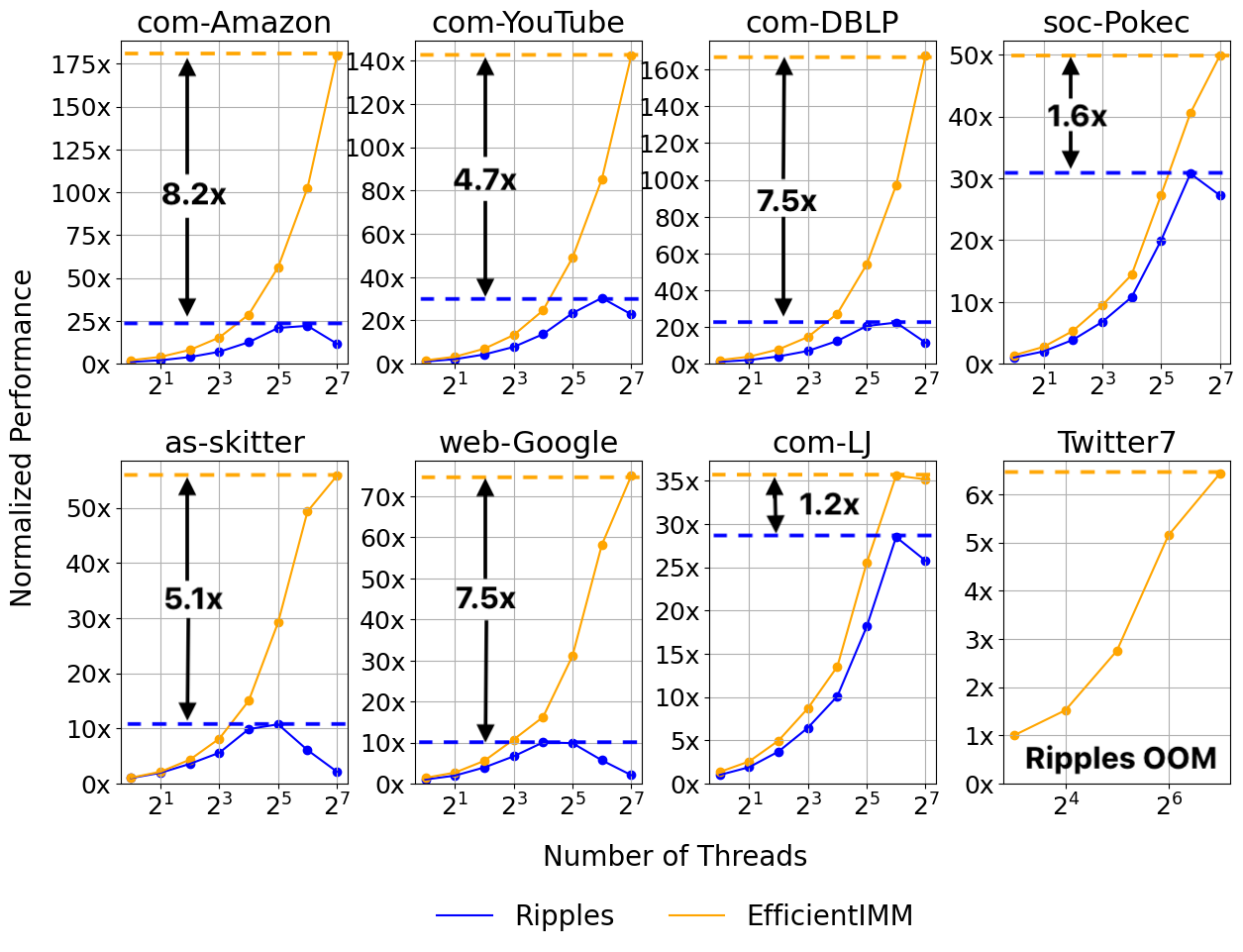}
  \caption{Strong Scaling Performance Normalized to 1 and 8 threads Ripples; IC Diffusion Model; k = 50, $\epsilon$ = 0.5}
  \label{fig:scaling_ic}
\end{figure}


\noindent
\textbf{Dataset Preparation:}
For our evaluation, we selected eight diverse graphs from the SNAP dataset~\cite{jure2014snap}. These graphs represent a variety of real-world networks, differing in size, complexity, and type (both directed and undirected), as detailed in Table~\ref{tab:merged_graph_rrrset}.
Based on these diverse collection of networks, we simulate the IC diffusion model by assigning uniformly random $[0, 1]$ edge probabilities. In the linear threshold (LT) diffusion model, weights are adjusted so that the probabilities of either activating a neighbor or activating none sum to one. 


\subsection{Performance Analysis}



\subsubsection{Runtime Performance and Strong Scaling} With the experimental setup, Table \ref{table:perf_comparison} shows the runtime performance in seconds. In strong scaling analysis, \sys consistently outperforms Ripples across SNAP datasets, achieving speedups ranging from 1.2x to 12.1x as shown in Figure \ref{fig:scaling_ic} and Figure \ref{fig:scaling_lt} for both IC and LT models. This improvement stems from \sys's superior scalability, capable of utilizing up to 128 threads, while Ripples struggles beyond 64 threads and, in some cases, is limited to as few as 2 threads. Key factors include \sys's efficient \textit{Find\_Most\_Influential\_Set} optimizations and adaptive RRRset representation, enabling it to run large datasets like Twitter7 without OOM. While \sys's performance slightly decreases on smaller datasets at 128 threads due to limited parallelization opportunities in small datasets, its overall scalability remains unmatched, particularly evident in its ability to handle larger graphs effectively. For Twitter7 graph, we run the experiments from 8 to 128 cores because Perlmutter allows at most 24 hours per workload. 

\subsubsection{Memory Traversal Evaluation}
\begin{table}[H]
  \centering
  \caption{Cache Misses of \sys v.s. Ripples in \textit{Find\_Most\_Influential\_Set} kernel}
  \centering
  \begin{tabular}{@{}l@{\hspace{3pt}}c@{\hspace{5pt}}c@{\hspace{5pt}}c@{}}
    \toprule
    \textbf{Graph} & \textbf{Ripples} & \textbf{EfficientIMM} & \textbf{Reduction} \\ & \textbf{(L1+L2 Misses)} & \textbf{(L1+L2 Misses)} \\
    \midrule
    com-Amazon     & 283,963,507  & 10,947,324 & 25.94x \\
    com-Google      & 406,351,077 & 18,139,797  & 22.40x \\
    soc-Pokec       & 48,114,540 & 516,602  & 93.14x \\
    com-YouTube & 135,802,513  & 379,979  & 357.39x \\
    com-LJ & 69,299,959  & 687,345  & 100.82x \\
    \bottomrule
  \end{tabular}
  \label{tab:memory_reduction}
\end{table}
\vspace{-5pt}
As discussed in section \ref{sec:proposed_algo}, the novel partitioning strategy employed by \sys significantly enhances the scalability of the \textit{Find\_Most\_Influential\_Set} algorithm, resulting in substantial speedups compared to the original Ripples framework. 
In here, we clarify how \sys's parallelization effectively reduces the number of load and store operations. In the referenced Ripples framework, vertices are partitioned across all the threads and each thread needs to search if the RRRsets contain the assigned vertices. During this phase, the memory overhead is quantified as $O(\log(\text{Average\_RRRsets\_Size}) \times \theta \times \text{Thread\_Count})$ since each thread employs binary search to ascertain the lower and upper bounds of the search endpoints within each RRRset.
In contrast, \sys distributes the RRRsets among the threads, obviating the need for redundant traversal and bounds determination checks. 
It directly conducts the counter update operations on the RRRsets and enhances efficiency. In the older CPU models, all the cores share the same memory and cache space and thus, the repetitive memory loads and stores will be alleviated by the more frequent cache hit. However, in our setup, the CPU used has 64 cores CPU and 4 NUMA nodes, so each NUMA node will have its own memory controller redundantly accessed the needed memory across NUMA nodes. 
With the combined software and system design, Table \ref{tab:memory_reduction} shows the profiled L1 + L2 cache misses between \sys and Ripples implementations, to reflect that \sys has better memory access pattern. On the 5 tested datasets,\sys results in significantly fewer cache misses, ranging from 22.4× in the com-Google dataset to 357.39× in Com-LJ.


\section{Related Work}
Our work optimizes sketch-based Influence Maximization (IM) algorithms for shared-memory environments. IMM, a state-of-the-art sketch-based algorithm, generates memory-intensive Reverse Influence Sketches (RIS) for influence sampling. HBMax reduces this memory usage by compressing sketches using Huffman encoding~\cite{chen2022hbmax}, but introduces codec overhead. In contrast, our approach enhances shared-memory performance and work efficiency without the overhead.

Modifying IMM's underlying mathematics allows specialization for different settings. For example, OPIM introduced an online algorithm based on IMM, enabling early termination of sampling when influence coverage is sufficient~\cite{tang2018OPIM}, which improves performance in resource-constrained scenarios.

PacIM is a shared-memory IM framework that constructs Forward Influence Sketches (FIS) for the Independent Cascade (IC) diffusion model~\cite{wang2024pacim}. Unlike IMM, which captures  \textit{Who Am I Influenced}, PacIM focuses on  \textit{Who I Am Influencing}. By memoizing activated nodes within connected components and leveraging novel parallel data structures, it achieves state-of-the-art performance for the IC model.

While our work concentrates on shared-memory optimization, it can be extended to distributed memory settings using MPI. Since our approach doesn't introduce additional communication compared to Ripples' MPI implementation, exploring an MPI extension is a promising direction for future work.

\section{Conclusion}

In this paper, we introduced \sys, a highly optimized framework for IMM that significantly enhances scalability and performance through advanced parallel processing strategies. 
\sys optimizes memory utilization, employs balanced parallelization strategies, and reduces synchronization overhead, effectively tackling the key challenges of scalability and efficiency in complex social networks.
Without sacrificing the accuracy, 
\sys consistently outperforms the existing Ripples framework, achieving speedups ranging from 1.2x to 12.3x and also reducing the L1+L2 cache misses by 22.4x to 357.39x in SNAP datasets.

\section*{Acknowledgement}
This research is based upon work supported by the Office of the Director of National Intelligence (ODNI), Intelligence Advanced Research Projects Activity (IARPA), through the Advanced Graphical Intelligence Logical Computing Environment (AGILE) research program, under Army Research Office (ARO) contract number $<W911NF22C0081>$. The views and conclusions contained herein are those of the authors and should not be interpreted as necessarily representing the official policies or endorsements, either expressed or implied, of the ODNI, IARPA, ARO, or the U.S. Government. We thank Pacific Northwest National Laboratory (PNNL), especially Marco Minutoli for the discussion related to work in the AGILE program.

\bibliographystyle{plain} 
\bibliography{references} 

\begin{thebibliography}{10}

\bibitem{journals/BroderKMRRSTW00}
Andrei~Z. Broder, Ravi Kumar, Farzin Maghoul, Prabhakar Raghavan, Sridhar Rajagopalan, Raymie Stata, Andrew Tomkins, and Janet~L. Wiener.
\newblock Graph structure in the web.
\newblock {\em Comput. Networks}, 33(1-6):309--320, 2000.

\bibitem{chen2022hbmax}
Xinyu Chen, Marco Minutoli, Jiannan Tian, Mahantesh Halappanavar, Ananth Kalyanaraman, and Dingwen Tao.
\newblock Hbmax: Optimizing memory efficiency for parallel influence maximization on multicore architectures.
\newblock In {\em Proceedings of the International Conference on Parallel Architectures and Compilation Techniques}, pages 412--425, 2022.

\bibitem{domingos2001mining}
Pedro Domingos and Matt Richardson.
\newblock Mining the network value of customers.
\newblock In {\em Proceedings of the seventh ACM SIGKDD international conference on Knowledge discovery and data mining}, pages 57--66, 2001.

\bibitem{jure2014snap}
Leskovec Jure.
\newblock Snap datasets: Stanford large network dataset collection.
\newblock {\em Retrieved December 2021 from http://snap. stanford. edu/data}, 2014.

\bibitem{kempe2003maximizing}
David Kempe, Jon Kleinberg, and {\'E}va Tardos.
\newblock Maximizing the spread of influence through a social network.
\newblock In {\em Proceedings of the ninth ACM SIGKDD international conference on Knowledge discovery and data mining}, pages 137--146, 2003.

\bibitem{leskovec2007celf}
Jure Leskovec, Andreas Krause, Carlos Guestrin, Christos Faloutsos, Jeanne VanBriesen, and Natalie Glance.
\newblock Cost-effective outbreak detection in networks.
\newblock KDD '07, page 420–429, New York, NY, USA, 2007. Association for Computing Machinery.

\bibitem{li2018survey}
Yuchen Li, Ju~Fan, Yanhao Wang, and Kian-Lee Tan.
\newblock Influence maximization on social graphs: A survey.
\newblock {\em IEEE Transactions on Knowledge and Data Engineering}, 30(10):1852--1872, 2018.

\bibitem{minutoli2020curipples}
Marco Minutoli, Maurizio Drocco, Mahantesh Halappanavar, Antonino Tumeo, and Ananth Kalyanaraman.
\newblock curipples: Influence maximization on multi-gpu systems.
\newblock In {\em Proceedings of the 34th ACM international conference on supercomputing}, pages 1--11, 2020.

\bibitem{minutoli2019fast}
Marco Minutoli, Mahantesh Halappanavar, Ananth Kalyanaraman, Arun Sathanur, Ryan Mcclure, and Jason McDermott.
\newblock Fast and scalable implementations of influence maximization algorithms.
\newblock In {\em 2019 IEEE International Conference on Cluster Computing (CLUSTER)}, pages 1--12. IEEE, 2019.

\bibitem{nersc_perlmutter}
{National Energy Research Scientific Computing Center}.
\newblock Perlmutter architecture documentation.
\newblock \url{https://docs.nersc.gov/systems/perlmutter/architecture/}, 2024.
\newblock Accessed: August 09, 2024.

\bibitem{shahrouz2021gim}
Soheil Shahrouz, Saber Salehkaleybar, and Matin Hashemi.
\newblock gim: Gpu accelerated ris-based influence maximization algorithm.
\newblock {\em IEEE Transactions on Parallel and Distributed Systems}, 32(10):2386--2399, 2021.

\bibitem{tang2018OPIM}
Jing Tang, Xueyan Tang, Xiaokui Xiao, and Junsong Yuan.
\newblock Online processing algorithms for influence maximization.
\newblock In {\em Proceedings of the 2018 International Conference on Management of Data}, SIGMOD '18, page 991–1005, New York, NY, USA, 2018. Association for Computing Machinery.

\bibitem{tang2015influence}
Youze Tang, Yanchen Shi, and Xiaokui Xiao.
\newblock Influence maximization in near-linear time: A martingale approach.
\newblock In {\em Proceedings of the 2015 ACM SIGMOD international conference on management of data}, pages 1539--1554, 2015.

\bibitem{wang2024pacim}
Letong Wang, Xiangyun Ding, Yan Gu, and Yihan Sun.
\newblock Fast and space-efficient parallel algorithms for influence maximization, 2024.

\bibitem{ye2012recomm}
Mao Ye, Xingjie Liu, and Wang-Chien Lee.
\newblock Exploring social influence for recommendation: a generative model approach.
\newblock SIGIR '12, page 671–680, New York, NY, USA, 2012. Association for Computing Machinery.

\bibitem{Zhang-numa-aware2015}
Kaiyuan Zhang, Rong Chen, and Haibo Chen.
\newblock Numa-aware graph-structured analytics.
\newblock PPoPP 2015, page 183–193, New York, NY, USA, 2015. Association for Computing Machinery.

\end{thebibliography}

\clearpage

\section*{Artifact Description/Artifact Evaluation}

\subsection{Artifact Meta Information}

{\small
\begin{itemize}
  \item \textbf{Algorithm}: Influence Maximization
  \item \textbf{Program}: EfficientIMM
  \item \textbf{Compilation}: C++17
  \item \textbf{Binary}: Binary not included
  \item \textbf{Dataset}: SNAP datasets, generated weights for IC and LT diffusion models
  \item \textbf{Run-time environment}: Developed and tested on Linux environment. Main software dependency is Anaconda and numactl
  \item \textbf{Hardware}: Developed and tested on Perlmutter, dual-socket AMD EPYC 7713 64-Core Processor
  \item \textbf{Output}: The experiments generate JSON files containing execution logs and the top \( k \) influential nodes (seed sets), stored in the \texttt{strong-scaling-logs-*} directories. Running \texttt{extract\_results.py} generates a summary performance comparison.
  \item \textbf{Publicly available?}: Yes
\end{itemize}
}

\subsection{Artifact Identification}

\subsubsection*{Contributions}

\begin{itemize}
  \item \textbf{C1}: Development of EfficientIMM, an optimized implementation of influence maximization algorithms based on the IMM algorithm, enhancing performance for computing influence maximization on large-scale networks using the Independent Cascade (IC) and Linear Threshold (LT) diffusion models.
  \item \textbf{C2}: Demonstration of significant performance improvements of EfficientIMM, especially on multi-NUMA systems over the original Ripples implementation when processing SNAP datasets, through comprehensive experimental evaluation.
\end{itemize}

\subsubsection*{Artifacts}

\begin{itemize}
  \item \textbf{A1}: EfficientIMM source code, including build scripts, experiment scripts, and analysis scripts to compile, run, and analyze EfficientIMM and Ripples, built on top of the Ripples framework (commit version \texttt{0a9f3e7c}), delivered through GitHub at \url{https://github.com/hxu296/EfficientIMM}.
  \item \textbf{A2}: SNAP datasets with generated weights for IC and LT diffusion models, delivered through Google Drive at \url{https://drive.google.com/file/d/1CRNC2NjSQ5B1\_Jngbg\_G4uCZzgWbG83Q/view?usp=sharing}.
\end{itemize}

These artifacts provide the necessary code, scripts, and datasets to reproduce the development and experimental evaluation of EfficientIMM, thereby substantiating \textbf{C1} and \textbf{C2}.

\subsection{Artifact Description}

\subsubsection{Software Dependencies}

Our artifact has been developed and tested on a Linux environment. The main software dependency is Anaconda. Other dependencies will be installed using it. We also require numactl to enable NUMA interleaving.

\subsubsection{Hardware Dependencies}

Our artifact has been developed and tested on Perlmutter, a dual-socket system with two AMD EPYC 7713 64-Core CPUs and 8 NUMA nodes in total. Similar hardware should result in comparable speedup results. It is recommended to run the experiments on a machine with at least 64 cores and 128\,GB of memory.

\subsubsection{Datasets}

We use the SNAP datasets with generated weights for the IC and LT diffusion models, available at:

\url{https://drive.google.com/file/d/1CRNC2NjSQ5B1\_Jngbg\_G4uCZzgWbG83Q/view?usp=sharing}

To programmatically download the datasets, run the provided script after setting up a dedicated \texttt{conda} environment.

\textit{Note}: The Twitter 7 dataset is not included due to computational constraints. All other datasets in paper are included.

\subsubsection{Build Instructions} First, clone the codebase:

\texttt{git clone \url{https://github.com/hxu296/EfficientIMM.git}}

\texttt{cd EfficientIMM}

Then, set up a dedicated \texttt{conda} environment:

\texttt{conda create -n efficientimm python=3.9}

\texttt{conda activate efficientimm}

To download and setup the datasets, run:

\texttt{bash download\_dataset.sh}

We use \texttt{conan} to automate packages installation, run:

\texttt{bash setup\_conan.sh}

Finally, build EfficientIMM and Ripples by executing:

\texttt{bash run\_build.sh}

This will build EfficientIMM and Ripples binaries with -O3.

\subsubsection{Experiment Execution Instructions}

To run the EfficientIMM and Ripples experiments on the SNAP datasets, execute:

\texttt{bash run\_efficient\_imm.sh}

\texttt{bash run\_ripples.sh}

These commands will run scaling experiments starting with 4 threads and doubling the thread count until the system limit is reached. All experiments use \( k = 50 \) and \( \epsilon = 0.5 \). The experiments may take about 5 hours to complete.

\subsection{Expected Results}

After the experiments finish, new directories will be created:

\begin{itemize}
  \item \texttt{strong-scaling-logs-ic-eimm}
  \item \texttt{strong-scaling-logs-lt-eimm}
  \item \texttt{strong-scaling-logs-ic-ripples}
  \item \texttt{strong-scaling-logs-lt-ripples}
\end{itemize}

Each JSON file in these directories contains the raw experiment results. To summarize and compare results, run:

\texttt{python3 extract\_results.py}

Two CSV files will be generated in the \texttt{results} directory: \texttt{speedup\_ic.csv} and \texttt{speedup\_lt.csv}. These files provide a human-readable speedup summary, allowing for an easy comparison between EfficientIMM and Ripples. The results should be similar to Table \ref{table:perf_comparison}, Fig \ref{fig:scaling_lt} and Fig \ref{fig:scaling_ic}. Each CSV file will contain the following columns:

\begin{itemize} 
\item \texttt{Dataset}
\item \texttt{Speedup}
\item \texttt{EfficientIMM Time (s)}
\item \texttt{Ripples Time (s)}
\item \texttt{Ripples Best \#Threads}
\item \texttt{EfficientIMM Best \#Threads}
\end{itemize}

To match the exact results, we suggest to run the aforementioned experiments on Perlmutter or the system with the same hardware. For the other systems, the expected results may vary and the best projected performance will be shown on a multi-NUMA system.

\end{document}